\documentstyle[prb,aps]{revtex}
\begin{document}
\title
{ BaCu${}_2$Si${}_2$O${}_7$: a new quasi 1-dimensional $S$ = 1/2
antiferromagnetic chain system}
\author{I. Tsukada, Y. Sasago,\cite[]{SasagoP} and K. Uchinokura}
\address{ Department of Applied Physics, The University of Tokyo, 7-3-1 Hongo, Bunkyo-ku, Tokyo 113-8656, Japan}
\author{A. Zheludev, S. Maslov, and G. Shirane}
\address{Physics Department, Brookhaven National Laboratory, Upton NY 11973-5000, USA}
\author{K. Kakurai}
\address{Neutron Scattering Laboratory, Institute for Solid State Physics, The University of Tokyo, Tokai,
Ibaraki 319-1106, Japan}
\author{E. Ressouche}
\address{MDN/SPSMS/DRFMC CEN Grenoble, 17 rue des Martyrs, 38054 Grenoble Cedex, France}

\date{\today}
\maketitle

\begin{abstract}
The new antiferromagnetic (AF) compound  BaCu${}_2$Si${}_2$O${}_7$ is studied
by magnetic susceptibility and neutron scattering techniques. The observed
behavior is  dominated by the presence of loosely coupled $S=1/2$ chains with 
the intrachain AF exchange constant $J_{\|}\approx 24.1$~meV. Long-range 
N\'{e}el ordering is observed below $T_{N}$ = 9.2~K. The results are discussed 
within the framework of the Mean Field- RPA model for weakly interacting 
quantum spin chains.
\end{abstract}


\section{Introduction} The amazing structural diversity of copper-oxide compounds makes
these materials very useful as model systems for fundamental studies of
low-dimensional magnetism. In most cases their unique properties result from
the particular topologies of the corresponding spin networks formed by
magnetic Cu${}^{2+}$ ions. Examples of such networks are found in CuGeO${}_3$
(linear chain: spin-Peierls), \cite{Hase1} CaCuGe${}_2$O${}_6$ (isolated
dimer), \cite{Sasago1,Zheludev1} BaCuSi${}_2$O${}_6$ (two-dimensional bilayer),
\cite{Sasago2} Sr${}_2$CuO${}_3$ (rigid linear chain), \cite{Ami1}
SrCu${}_2$O${}_3$ (two-leg ladder), \cite{Hiroi1} and SrCuO${}_2$ (edge-shared
double linear chain), \cite{Motoyama1} {\it etc}. Magnetic interactions in all
these systems are determined by the microscopic spatial coordination of
Cu${}^{2+}$ and O${}^{2-}$.\cite{Mizuno1} The so-called corner-sharing
($\angle$Cu-O-Cu $\approx$ 180$^{\circ}$) and edge-sharing ($\angle$Cu-O-Cu
$\approx$ 90$^{\circ}$) configurations have been extensively studied. A far
richer spectrum of properties is expected from systems with intermediate bond
angles, and the search for realizations of such coupling geometries has become
important.

The rather poorly studied BaCu${}_2$Si${}_2$O${}_7$ is a good example of a
zig-zag chain network of corner-sharing CuO$_4$ plaquettes  (Fig.~\ref{Fig.1})
with intermediate values of $\angle$Cu-O-Cu bond angles, 124${}^{\circ}$. The
cystal structure is orthorhombic, space group $Pnma$, with the cell constants 
being $a=6.862(2)$~{\AA}, $b=13.178(1)$~{\AA}, and $c=6.897(1)$~{\AA}.
\cite{Oliveira1} The spin chains run along the $c$ crystallographic axis. In 
the present paper we report magnetic susceptibility measurements and inelastic 
neutron scattering experiments on this material. We find it to be an excellent 
model of weakly-coupled quantum $S=1/2$ chains, exhibiting a crossover from 1-D
behavior at high temperatures to 3-D behavior and long-range AF (N\'eel) order 
at low temperatures. As anticipated for a system with intermediate values of 
bond angles, the magnetic properties are extremely sensitive to slight 
modifications in atomic positions. This is demonstrated by preliminary 
susceptibility results for BaCu${}_2$Ge${}_2$O${}_7$, which, unlike the 
isomorphous BaCu${}_2$Si${}_2$O${}_7$, orders in a weak-ferromagnetic 
structure with a small net magnetization.

\section{Experimental}
Polycrystalline samples of BaCu${}_2$Si${}_2$O${}_7$
(or Ba\-Cu${}_2$\-Ge${}_2$\-O${}_7$) were prepared by conventional solid-state
reaction method, using BaCO${}_3$, SiO${}_2$ (or GeO${}_2$), and CuO as
starting materials. The polycrystalline rods were crushed to fine powder for 
neutron powder diffraction studies. Single crystals were grown using the 
floating-zone technique from a sintered polycrystalline rod. In inelastic 
neutron scattering experiments we utilized an as-grown single-crystal
rod (5~mm diameter $\times$ 10~mm long). A small fragment cut to a 3~mm${}^3$
cube was used in single-crystal susceptibility measurement.

All magnetic susceptibility experiments were done with a commercial SQUID
magnetometer ($\chi$-MAG7, Conductus, Co. Ltd.) in the temperature range of 2
to 300~K. Neutron powder diffraction experiments were carried out at the 400
cells CRG diffractometer D1B at Institut Laue Langevin (Grenoble, France).
Single-crystal neutron scattering studies were performed with the ISSP-PONTA
triple-axis spectrometer installed at 5G beam hole of the Japan Research 
Reactor 3M at the Japan Atomic Energy Research Institute (Tokai, Japan). 
Pyrolytic graphite PG(002) reflections were used in the monochromator and 
analyzer. The collimation setup was 60'-80'-40'-80', with a PG filter 
positioned after the sample. The final neutron energy was fixed at $E_f$ = 
14.7~meV. The sample was mounted in a standard ``ILL Orange'' cryostat with the
(0, $k$, $l$) reciprocal-space plane in the scattering plane of the 
spectrometer, and the data were collected in the temperature range 2 - 15~K.

\section{Results}
\subsection{Magnetic susceptibility}
The temperature dependences of DC magnetic susceptibility measured under $H$ =
10~kOe in BaCu${}_2$Si${}_2$O${}_7$ and BaCu${}_2$Ge${}_2$O${}_7$ are shown in
Fig.~\ref{Fig.2}(a). The experimental curve for BaCu${}_2$Si${}_2$O${}_7$ shows
a sharp peak around 9~K and a broad maximum around 150~K. The
BaCu${}_2$Si${}_2$O${}_7$ data were fitted to the theoretical Bonner-Fisher 
(BF) curve for a one-dimensional $S=1/2$ quantum antiferromagnet\cite{Bonner1}
(solid lines). A Heisenberg exchange constant $J \approx$ 280~K (24.1~meV) was
obtained from this analysis.\cite{defJ} Note that the BF fit becomes rather 
poor in the low-temperature region, suggesting the onset of 3-dimensional spin
correlations in this regime.

To identify the low-temperature ordered structure of BaCu${}_2$Si${}_2$O${}_7$
below 9.2~K, the anisotropy of magnetic susceptibility was studied in a 
single-crystal sample. The measured temperature dependences are shown in
Fig.~\ref{Fig.2}(b). Signs of long-range N\'eel ordering are clearly seen.
Below 9.2~K a substantial drop is observed in $\chi_c$, while both $\chi_a$ and
$\chi_b$ change little from 9.2~K down to 2~K. This clearly points to the
presence of antiferromagnetic long-range order below $T_N$ = 9.2~K with
crystallographic $c$ axis being the magnetic easy axis of the system. Note that
at $T\rightarrow 0$ $\chi_c$ retains a substantial non-zero value, suggesting a
reduction of ordered moment, presumably due to the 1-D nature of
the system. As expected, the one-dimensional character of paramagnetic phase is
observed in single crystals as well. The susceptibility shows broad maxima 
around 180~K along all the three crystallographic directions. Fitting these 
data to a BF curve, and assuming an anisotropic $g$-factor, we obtain $g_a$ = 
2.5 and $g_b=g_c$ = 2.2.

For BaCu${}_2$Ge${}_2$O${}_7$ a BF analysis (dashed line in
Fig.~\ref{Fig.2}(a)) of the high-$T$ part of the experimental $\chi(T)$ curve
yields $J\approx 540$~K (46.5 meV), i.e., much larger than for
Ba\-Cu${}_2$\-Si${}_2$\-O${}_7$. In the low-temperature region the magnetic
susceptibilities of the two isomorphous compounds become qualitatively
different. The drastic increase of magnetization in BaCu${}_2$Ge${}_2$O${}_7$
observed below 8.5~K is attributed to the appearance of uniform spontaneous
magnetization. This increase is particularly well seen in the low-field
measurement (Fig.~\ref{Fig.2}(a), inset). The saturation value of magnetzation
($\approx 6\times 10^{-2}$~emu/g) is orders of magnitude less than 22.15
emu/g, that one would obtain if all the Cu$^{2+}$ spins were aligned
ferromagnetically. The clear dominance of $c$-axis antiferromagnetism in the
paramagnetic phase  of BaCu${}_2$Ge${}_2$O${}_7$ suggests that the spontaneous
uniform moment below 8.5~K is a result of a canted weak-ferromagnetic spin
arrangement.

\subsection{Neutron diffraction.}
As a first step towards determining the magnetic structure of
BaCu$_{2}$Si$_{2}$O$_{7}$ we performed neutron powder diffraction experiments
at 2~K and 15~K, i.e., above and below the N\'eel temperature. Subtraction of
the powder scan data collected at these two temperatures did not reveal any
sign of magnetic Bragg reflections. This failure to observe any magnetic signal
in the powder experiment can be attributed to a small value of ordered moment
and the fact that all magnetic reflections appear on top of strong nuclear
peaks (see below), not surprising for a structure with 8 magnetic Cu$^{2+}$
ions per unit cell, and are thus very difficult to detect.

Even in the subsequent single-crystal experiment, only limited information  on
the magnetic structure of AF phase of BaCu${}_2$Si${}_2$O${}_7$ could be
obtained. Only one magnetic reflection, namely $(011)$, could be reliably
measured at low temperatures. The $(011)$ {\it nuclear} reflection is allowed
by crystal symmetry, but its intensity is fortunately relatively weak. The
inset (a) in Fig.~\ref{Fig.3} shows rocking curves of the $(011)$ Bragg peak
measured above (open circles) and below (solid circles) the magnetic transition
temperature. The difference between these two scans is shown in the inset (b) 
of Fig.~\ref{Fig.3}. The measured peak intensity is plotted against temperature
in the main panel of Fig.~\ref{Fig.3}. The observed increase upon cooling 
through $T_{N} = 9.2$~K is attributed to long-range magnetic ordering. 
Intensities of several other Bragg peaks, including $(031)$, $(013)$, $(051)$,
$(071)$, $(053)$, and $(035)$ were also measured as functions of temperature.
However, the nuclear contribution to these reflections is much larger than that for $(011)$, and no intensity increase could be observed beyond the 
experimental error bars upon cooling through $T_{N}$.

\subsection{Inelastic scattering}
Inelastic neutron scattering studies of spin excitations in
BaCu${}_2$Si${}_2$O${}_7$\ were also performed. The rather steep spin wave
dispersion along the chain axis was measured in two constant-$E$ scans near the
1-D AF zone-center $\vec{Q}=(0, -1.5, l)$, at energy transfers $\hbar \omega = 
10$~meV (Fig.~\ref{Fig.4}(a)) and $\vec{Q}=(0, 0, l)$ at $\hbar \omega=5$~meV 
(Fig.~\ref{Fig.4}(b)). Both data sets were collected at $T=2$~K, and transverse
wave vector $k$ was chosen to optimize focusing conditions. 
Even in the 10~meV scan the two spin wave branches are barely resolved.
Unfortunately, for higher energy transfers the magnetic signal was too weak to
be measured in the present experiment.

The spin wave dispersion along the $b$ axis was measured at $T=2$~K in
constant-$Q$ scans at three reciprocal-space points at the bottom of
longitudinal dispersion (Fig.~\ref{Fig.5}): $\vec{Q}=(0,0,1)$,
$\vec{Q}=(0,0.5,1)$, and $\vec{Q}=(0,1.1,1)$. Clearly the transverse bandwidth
($\approx 1$~meV) is very small compared to that along the chain axis, which
confirms the 1-D character of spin correlations in BaCu${}_2$Si${}_2$O${}_7$.
Even in the vicinity of the 3-D magnetic zone-center $(011)$ the excitation
energy appears to extrapolate to a non-zero value, i.e., the spectrum has a gap
$\Delta_{(011)}\approx1.5$~meV.

\section{Discussion}
All our experimental results point to that both BaCu${}_2$Si${}_2$O${}_7$ and
BaCu${}_2$Ge${}_2$O${}_7$ should be considered as quasi-1D systems, dominated
by strong intrachain antiferromagnetic exchange interactions. Long-range
ordering at low temperatures occurs owing to a much weaker interchain coupling.
The subtleties of the low-temperature magnetic structures and magnetic
anisotropy, as well as the variations of the intrachain coupling constant $J$
in the two systems are expected to be related to the details of the microscopic
atomic arrangement.

\subsection{Microscopic structure and magnetic interactions}
Rather remarkable is the drastic (a factor of 2) difference in the intrachain
AF exchange constants of  the two isomorphous systems BaCu${}_2$Si${}_2$O${}_7$
and BaCu${}_2$Ge${}_2$O${}_7$. The AF spin chains in these species are formed
by corner-sharing CuO$_4$ plaquettes, and the main contribution to AF
interactions is expected to be the standard superexchange mechanism involving
the shared O sites. A single chain is visualized in Fig.~\ref{Fig.6}(a).
Only the Cu sites and the shared oxygen sites are shown. The Cu$^{2+}$ ions
are almost perfectly lined up. However, the Cu-O-Cu bond angle is
substantially smaller than $180^{\circ}$. The difference in $J$ is most likely
due to a difference in this crucial bonding angle: $124^{\circ}$ in
BaCu${}_2$Si${}_2$O${}_7$ and $135^{\circ}$ in the Ge-based system.
\cite{Oliveira1} Indeed we know that a larger bond angle is more favorable
for superexchange involving oxygen.

Subtle structural differences between the two systems are also to be held 
responsible for the difference in the low-$T$ behavior. In addition to
single-ion anisotropy that manifests itself in the anisotropy of the $g$-factor
and the gap in the spin wave spectrum (see discussion below), one should also
consider the possibility of Dzyaloshinskii-Moriya asymmetric exchange
interaction. The latter is allowed by local Cu-site symmetry in
BaCu${}_2$Si${}_2$O${}_7$ and BaCu${}_2$Ge${}_2$O${}_7$, and may be responsible
for the weak ferromagnetism of BaCu${}_2$Ge${}_2$O${}_7$. The correlation
between the microscopic structure and magnetic properties  of
BaCu${}_2$Si${}_2$O${}_7$ and BaCu${}_2$Ge${}_2$O${}_7$ is indeed a very
interesting problem. However, at the present stage we do not have sufficient
information required for a more detailed discussion of this subject, and
suggest that further experiments, possibly using polarized neutron diffraction,
should be performed to clarify the magnetic structures in the ordered state.

\subsection{Weakly coupled quantum spin chains}

Ignoring for now such subtleties as magnetic anisotropy, spin canting in the
ordered state and the possibility of DM-interaction, we shall turn to
discussing what makes the two new systems really valuable for our cause, namely
the 1-D quantum antiferromagnetic aspect of their magnetic properties. In the
following sections we shall demonstrate that the observed behavior of
BaCu${}_2$Si${}_2$O${}_7$ can be very well understood within the framework of
existing theories for weakly-coupled $S=1/2$ AF chains. Particularly useful for
describing such systems are the chain-MF (Mean Field)
\cite{Scalapino75,Affleck94,Schulz96} and the corresponding chain-RPA (Random
Phase Approximation) models.\cite{Schulz96,Essler1} These approaches were
previously shown to work extremely well for a number of well-characterized
compounds (see for example Refs.~\onlinecite{Tennant95,Kojima97} and
references therein).

\subsubsection{A model Hamiltonian}
Before we can apply these MF-RPA theories to our particular system, we have to
construct a model spin Hamiltonian for BaCu${}_2$Si${}_2$O${}_7$. A slight
complication arises from a non-Bravais arrangement of magnetic ions in the
crystal. Indeed, the fractional cell coordinates of Cu$^{2+}$ are 1)
$(0.25-\delta _{x},\delta_{y},0.75+\delta
_{z})$, 2) $(0.25+\delta _{x},-\delta _{y},0.25+\delta _{z})$, 3) $%
(0.75-\delta _{x},0.5-\delta _{y},0.75-\delta _{z})$, 4) $(0.75+\delta
_{x},0.5+\delta _{y},0.25-\delta _{z})$, 5) $(0.75+\delta _{x},-\delta
_{y},0.25-\delta _{z})$, 6) $(0.75-\delta _{x},\delta _{y},0.75-\delta _{z})$%
, 7) $(0.25+\delta _{x},0.5+\delta _{y},0.25+\delta _{z})$, and 8) $%
(0.25-\delta _{x},0.5-\delta _{y},0.75+\delta _{z})$, where $\delta
_{x}=0.028$, $\delta _{y}=0.004$, and $\delta _{z}=0.044$. The shifts $\delta
_{x}$ and $\delta _{z}$ present few problems, as they do not disturb the
equivalence of nearest-neighbor Cu-Cu bonds along the $a$ and $c$ axes,
respectively. $\delta _{y}$, on the other hand, leads to alternating
nearest-neighbor Cu-Cu distances along the $b$ axis. At present we shall ignore
this slight alternation, assuming $\delta _{y}\approx 0$ and postulating
nearest-neighbor Cu-Cu bonds to be equivalent along the $b$ axis as well. The
simplest nearest-neighbor Heisenberg Hamiltonian can then be written as
\begin{equation}
{\hat H}=\sum_{m,n,p}{\bf S}_{m,n,p}\left[ J_{x}{\bf S}_{m+1,n,p}+J_{y}{\bf S}%
_{m,n+1,p}+J_{z}{\bf S}_{m,n,p+1}\right]\label{ham}
\end{equation}
Here $m$, $n$, and $p$ enumerate the spins ${\bf S}_{m,n,p}$ along the $a$,
$b$, and $c$ axes, respectively. As BaCu${}_2$Si${}_2$O${}_7$\ is clearly a
quasi 1-D antiferromagnet, $J_{z}>0$ and $|J_{z}|\gg |J_{y}|, |J_{x}|$.

\subsubsection{Reduction of ordered moment}

One of the most important predictions of the chain-MF model is the relation
between ordering temperature, magnitude of interchain coupling, and saturation
moment at $T\rightarrow 0$. At the MF level the actual signs of magnetic
interactions are not important and these properties are determined by the mean
interchain coupling constant $|J_{\perp}|=(|J_x|+|J_y|) / 2$. Schulz gives the 
relation between $m_0$, $T_N$ and $|J_{\perp}|$:\cite{Schulz96}

\begin{equation}
|J_{\bot}|=\frac{T_{N}}{1.28\sqrt{\ln(5.8J/T_{N})}},
\label{Jperp}
\end{equation}
\begin{equation}
m_0\approx1.017\sqrt{\frac{|J_{\perp}|}{J}}.
\label{moment}
\end{equation}
In our case of BaCu${}_2$Si${}_2$O${}_7$ the susceptibility data ($T_N$ =
0.793~meV and $J$ = 24.1~meV) gives $m_0=0.11\mu_{B}$.

To test this theoretical prediction we have to determine the spin structure
in the ordered state. Unfortunatelly, with only one measured magnetic
reflection we can only speculate on this. Let us consider an approximate 
collinear magnetic state, with all spins strictly along $(001)$, as suggested
by $\chi(T)$ measurements. The two simplest possible AF structures (A and B) 
consistent with the presence of a substantial $(011)$ magnetic peak are shown 
in Figs.~\ref{Fig.7}(a) and (b). We can estimate the magnitude of saturation 
moment that would be required in structures A and B to produce a $(011)$ 
magnetic peak of measured intensity. Comparing the measured magnetic intensity 
of $(011)$ to that of two strong nuclear reflections $(020)$ and $(022)$, and 
making use of the known room-temperature crystal structure,\cite{Oliveira1} for
the saturated moment $m_0$ of Cu$^{2+}$ we get $m_0=0.16\mu_{B}$ and 
$m_0=0.55\mu_{B}$ for structures A and B, respectively. A simple calculation of nuclear and magnetic structure factors indicates that a magnetic moment as 
large as $0.55\mu_{B}$ would be easily detected in a powder experiment at 
($h$, $k$, $l$-odd) positions. The lack of any clear magnetic signal in our 
powder data suggests that of the two proposed structures only structure A can 
be realized in BaCu${}_2$Si${}_2$O${}_7$. For structure A the estimated ordered
moment is in rather good agreement with the predictions of Eq.~(\ref{moment}).
We shall thus use structure A as a working assumption for the spin arrangement 
in BaCu${}_2$Si${}_2$O${}_7$. For the Hamiltonian (\ref{ham}) to stabilize this
type of ground state we will have to assume the exchange interactions to be
ferromagnetic along the $a$ axis ($J_x <$ 0), and antiferromagnetic along the 
$b$ axis ($J_y >$ 0).

\subsubsection{Spin dynamics}
As is well known, for an {\it isolated} antiferromagnetic $S=1/2$ chain the
dynamic structure factor can be described as a two-spinon excitation
continuum.\cite{Muller81} To a good approximation
\begin{equation}
S(q,\omega)\propto\frac{1}{\sqrt{\omega^2-\omega_{q}^2}}\Theta(\omega-\omega_{q})\label{e1},
\end{equation}
where $\omega_{q}$ is the lower bound of the continuum given by
\begin{equation}
\hbar \omega_{q}=\frac{\pi}{2} J_{\|}|\sin(q)|\label{e2},
\end{equation}
and $J_{\|}$ is intrachain coupling constant. This type of behavior has been
seen experimentally in a number of systems, in particular KCuF$_3$ (Refs. 
\onlinecite{Tennant95,Nagler91}) and CuGeO$_3$ (Ref. \onlinecite{Arai}). As 
mentined above, spin correlations in {\it weakly coupled}
quantum $S=1/2$ spin chains are very well described by the chain-RPA
model.\cite{Schulz96,Essler1} As the system becomes ordered in three
dimensions, the spectrum develops a mass gap $\Delta(\vec{Q}_{\bot})$ that
depends on the transverse momentum transfer $\vec{Q}_{\bot}$. The bandwidth of
transverse dispersion is of the order of $J_{\bot}$. This is in striking
contrast with classical spin wave theory, where the transverse bandwidth is
proportional to $\sqrt{J_{\bot}J_{\|}}$. The mass gap goes to zero only at the
3-D magnetic zone-centers, i.e., at the position of magnetic Bragg reflections.
A sharp single-magnon mode is the lowest-energy excitation that is split off
from the lower bound of the continuum. A two-magnon continuum then starts at
$2\Delta(\vec{Q}_{\bot})$. The dispersion of the magnon branch is given by
\begin{equation}
(\hbar \omega_{\vec{Q}})^2=
\frac{\pi^2}{4}J_{\|}^2\sin^2(Q_{\|})+\Delta^2(\vec{Q}_{\bot}).\label{harekrishna}
\end{equation}
For the interchain coupling geometry of KCuF$_3$ (equal nearest-neighbor
ferromagnetic interactions along the $a$ and $b$ axes) the expression for
$\Delta(\vec{Q}_{\bot})$  has been derived in Ref.~\onlinecite{Essler1}.

Near the bottom of 1-D dispersion the single mode approximation (SMA) works
very well\cite{Schulz96} and to a good approximation the dynamic structure
factor at $T=0$ is the given by
\begin{equation}
S(\vec{Q},\omega)\propto\frac{1}{\omega}\delta(\omega-\omega_{\vec{Q}})\label{sqw2}.
\end{equation}
For $\hbar\omega\gg |J_{x}|, |J_{y}|$, on the other hand, it is more
appropriate to use Eq.~(\ref{e1}) for isolated chains.

The dynamic structure factor $S(\vec{Q},\omega )$ for the non-Bravais spin
lattice in BaCu${}_2$Si${}_2$O${}_7$\ can be expressed through the dynamic
structure factor $S_{0}(\vec{Q},\omega )$ of an equivalent system with the same
exchange constants and a Bravais spin lattice. The latter is obtained by
setting $\delta _{x}$, $\delta _{y}$ and $\delta _{z}$ to zero. It is
straightforward to show that:
\begin{eqnarray}
S(\vec{Q},\omega ) =\cos ^{2}\left( 2\pi h\delta _{x}\right) \cos ^{2}\left(
2\pi l\delta _{z}\right) S_{0}(\vec{Q},\omega )\nonumber \\ +\cos ^{2}\left(
2\pi h\delta _{x}\right) \sin ^{2}\left( 2\pi l\delta _{z}\right) S_{0}\left(
\vec{Q}+\{001\},\omega \right)\nonumber \\ + \sin ^{2}\left( 2\pi h\delta
_{x}\right) \cos ^{2}\left( 2\pi l\delta _{z}\right) S_{0}\left(
\vec{Q}+\{100\},\omega \right)\nonumber \\ +\sin ^{2}\left( 2\pi h\delta
_{x}\right) \sin ^{2}\left( 2\pi l\delta _{z}\right) S_{0}\left(
\vec{Q}+\{101\},\omega \right).
\end{eqnarray}
Here we have defined $Q=(\frac{2\pi }{a}h,\frac{2\pi}{b}k,\frac{2\pi }{c}l)$.
In our particular case $\delta _{x}, \delta _{z}\ll 1$, so, for not too large
momentum transfers $S(\vec{Q},\omega )\approx S_{0}(\vec{Q},\omega )$. In other
words, we can safely analyze the measured inelastic scans assuming an idealized
Bravais arrangement of Cu$^{2+}$ sites.

We can now rewrite Eqs.~(\ref{e1}) and (\ref{e2}) to match the notation
introduced above for BaCu${}_2$Si${}_2$O${}_7$:
\begin{eqnarray}
S(\vec{Q},\omega)\propto
\frac{1}{\sqrt{\omega^2-\omega_{Q_{\|}}^2}}\Theta(\omega-\omega_{Q_{\|}})\label{sqw1},\\
\hbar \omega_{Q_{\|}}=\frac{\pi}{2} J_{z}|\sin(\pi l)|.
\end{eqnarray}
The result for transverse dispersion  $\Delta(\vec{Q}_{\bot})$ derived for the
case of KCuF$_3$ by Essler {\it et al}.\cite{Essler1} can also be easily 
adapted for use with the coupling geometry in BaCu${}_2$Si${}_2$O${}_7$:

\begin{equation}
 (\hbar \omega_{\vec{Q}})^2=
 \frac{\pi^2}{4}J_z^2\sin^2(\pi l)+A^2\frac{J_y-J_x}{4}
 \left[J_y-J_x+J_x\cos(\pi h)+J_y\cos(\pi k)\right]+D^2.
\end{equation}
In this formula we introduced an empirical anisotropy gap $D$. The
dimensionless mass gap $A$ is given by $A\approx6.175$.\cite{Schulz96}

\subsubsection{Analysis of inelastic data}
The dynamic structure factor, Eq.~(\ref{sqw1}), convoluted with the
4-dimensional spectrometer resolution function was used to analyze
the two measured constant-$E$ scans. Reasonably good fits to the
data (solid lines in Fig.~\ref{Fig.4}) were obtained with only three
adjustable parameters, namely $J_z$, an intensity prefactor, and a
flat background for each scan. The refined value
$J_{z}=19.84$~meV is in agreement with the previous estimation, 24.1~meV, 
based on the experimental $\chi(T)$ curve.

For analyzing the $b$-axis dispersion we used the SMA given by
Eq.~(\ref{sqw2}), also convoluted with the resolution function. 
As we do not have any data for the dispersion along the $a$ axis, to reduce the
number of parameters we have assumed $|J_x|=|J_y|=J_{\bot}$. The relevant 
adjustable parameters for the fit were thus $J_{\bot}$, the anisotropy constant
$D$, responsible for the gap at $(011)$, and an intensity prefactor. The 
fitting procedure gives $J_{\bot}=0.29(2)$~meV and $D=1.59(4)$~meV. The 
resulting simulated scans are shown in solid lines in Fig.~\ref{Fig.4}. On the 
other hand, $J_{\perp}$ is estimated from the susceptibility data using
Eq.~(\ref{Jperp}),\cite{Schulz96} which gives $J_{\perp}$ = 0.27~meV for 
BaCu${}_2$Si${}_2$O${}_7$, in excellent agreement with what we find from the
analysis of transverse dispersion.

\section{Concluding remarks}
The purpose of this paper was to introduce  BaCu${}_2$Si${}_2$O${}_7$ as a new
model $S=1/2$ quantum antiferromagnet. The particular combination of intrachain
and transverse coupling constants make future neutron scattering studies of
this material particularly promising. Indeed, BaCu${}_2$Si${}_2$O${}_7$ is a
{\it better} 1-D compound than KCuF$_3$ with a smaller ratio of $T_{N}$ / 
$J_{\|}$. At the same time the interchain interactions in the new silicate are
sufficiently strong to make the mass gap easily observable with inelastic
neutron scattering techniques. The use of cold neutrons should enable an
experimantal study of the double gap, i.e., the separation between the magnon
branch and the 2-particle continuum. This effect in BaCu${}_2$Si${}_2$O${}_7$
is expected be similar to the double gap found in the dimerized state of
CuGeO$_3$,\cite{Regnault97} but is caused by interchain interactions, rather
than dimerization within the chains.

When this work was in progress we became aware of the similar susceptibility
data of polycrystalline samples of BaCu${}_2$Si${}_2$O${}_7$ and
BaCu${}_2$Ge${}_2$O${}_7$. \cite{Yamada1} These results are consistent with our
data.

\acknowledgements We would like to thank T. Masuda, T. Yamada, and Z. Hiroi
for valuable discussions. We also thank K. Nakajima, and T. Yosihama for
technical assistance of neutron experiments. This work is supported in part by
the U.S. -Japan Cooperative Program on Neutron Scattering, the Grant-in-Aid
for Scientific Research on Priority Area ``Mott transition'', Grant-in-Aid for
COE Research ``SCP coupled systems'', and Grant-in-Aid for Scientific Research
(A) of the Ministry of Education, Science, Sports, and Culture. Work at
Brookhaven National Laboratory was carried out under Contract No.
DE-AC02-76CH00016, Division of Material Science, U.S.\ Department of Energy.

\begin{figure}
\caption{ A schematic view of the crystal structure of
BaCu${}_2$Si${}_2$O${}_7$. Chains of corner-sharing CuO$_4$ plaquettes run
along the $c$ crystallographic axis and are supported by a network of
SiO$_4$ tetrahedra.}
\label{Fig.1}
\end{figure}

\begin{figure}
\caption{ (a) Magnetic susceptibility of polycrystalline samples of
BaCu${}_2$Ge${}_2$O${}_7$ and BaCu${}_2$Si${}_2$O${}_7$ measured at $H$ =
10~kOe. Solid and dotted lines are fits to the theoretical Bonner-Fisher
curve.
The inset shows an evolution of magnetization in BaCu${}_2$Ge${}_2$O${}_7$, 
measured in a weak magnetic field of 100 Oe. (b) Magnetic susceptibility of 
single-crystal BaCu${}_2$Si${}_2$O${}_7$ measured in $H$ = 1~kOe magnetic 
field applied along the principal axes. Solid and dotted lines are as in (a). 
The inset shows magnified curves around $T_N$. }
\label{Fig.2}
\end{figure}

\begin{figure}
\caption{ Inset: (a) Elastic ${\theta}-2{\theta}$ neutron scans through the
(011)
Bragg reflection collected above and below the magnetic ordering temperature
in
BaCu${}_2$Si${}_2$O${}_7$. (b) The corresponding magnetic contribution
obtained
by subtracting the data shown in (a). Main panel: measured temperature
dependence of the (011) peak intensity. Below $T_N$ = 9.2~K the magnetic Bragg
peak emerges on top of the
(011) nuclear reflection.} \label{Fig.3}
\end{figure}

\begin{figure}
\caption{ Constant-energy scan along the $(0, 1.5, l)$ direction measured at
$T$ = 2 K in BaCu${}_2$Si${}_2$O${}_7$, for energy transfers (a) $\hbar
\omega=10$~meV, and (b) $\hbar \omega=5$~meV. Solid lines are theoretical
fits as described in the text. } \label{Fig.4}
\end{figure}

\begin{figure}
\caption{ Constant-$Q$ scans measured in BaCu${}_2$Si${}_2$O${}_7$ at $T$ =
2
K. (a) ${\bf Q}$ = (0, 0, 1), (b) ${\bf Q}$ = (0, 0.5, 1), and (c) ${\bf }$
= (
0, 1.1, 1). Solid lines are single-mode approximation fits, as described in
the
text. } \label{Fig.5}
\end{figure}

\begin{figure}
\caption{ A schematic representation of the magnetic ion arrangement in
BaCu${}_2$Si${}_2$O${}_7$. (a) A single Cu-O chain. Cu-O-Cu bond angles are
shown explicitly. (b) Relative arrangement of individual chains, projected
onto
the $(ab)$ crystallographic plane. Interchain interactions are characterized
by
two nearest-neighbor exchange constants $J_x$ and $J_y$. } \label{Fig.6}
\end{figure}

\begin{figure}
\caption{ Two simple collinear magnetic structures consistent with the
available data on BaCu${}_2$Si${}_2$O${}_7$, with ferromagnetic (A) and
antiferromagnetic (B) nearest-neighbor coupling along the $a$-axis. }
\label{Fig.7}
\end{figure}


\begin{references}
\bibitem{defJ} To avoid confusion, throughout the paper we will be using the 
notation for $J$ in which the 1-D Hamiltonian is written as 
${\hat H}=J{\sum}_i{\bf S}_i{\cdot}{\bf S}_{i+1}$.
\end{references}
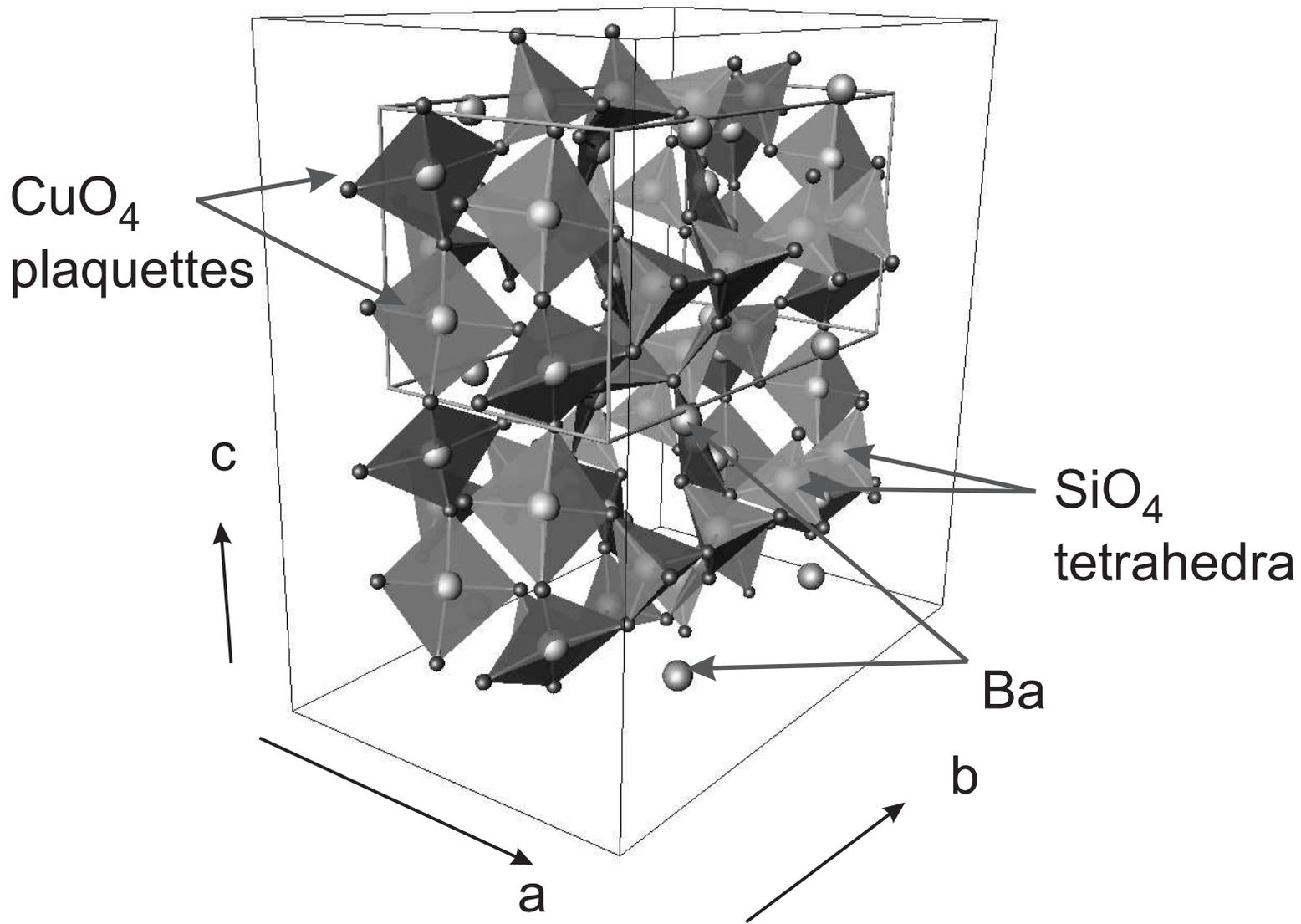
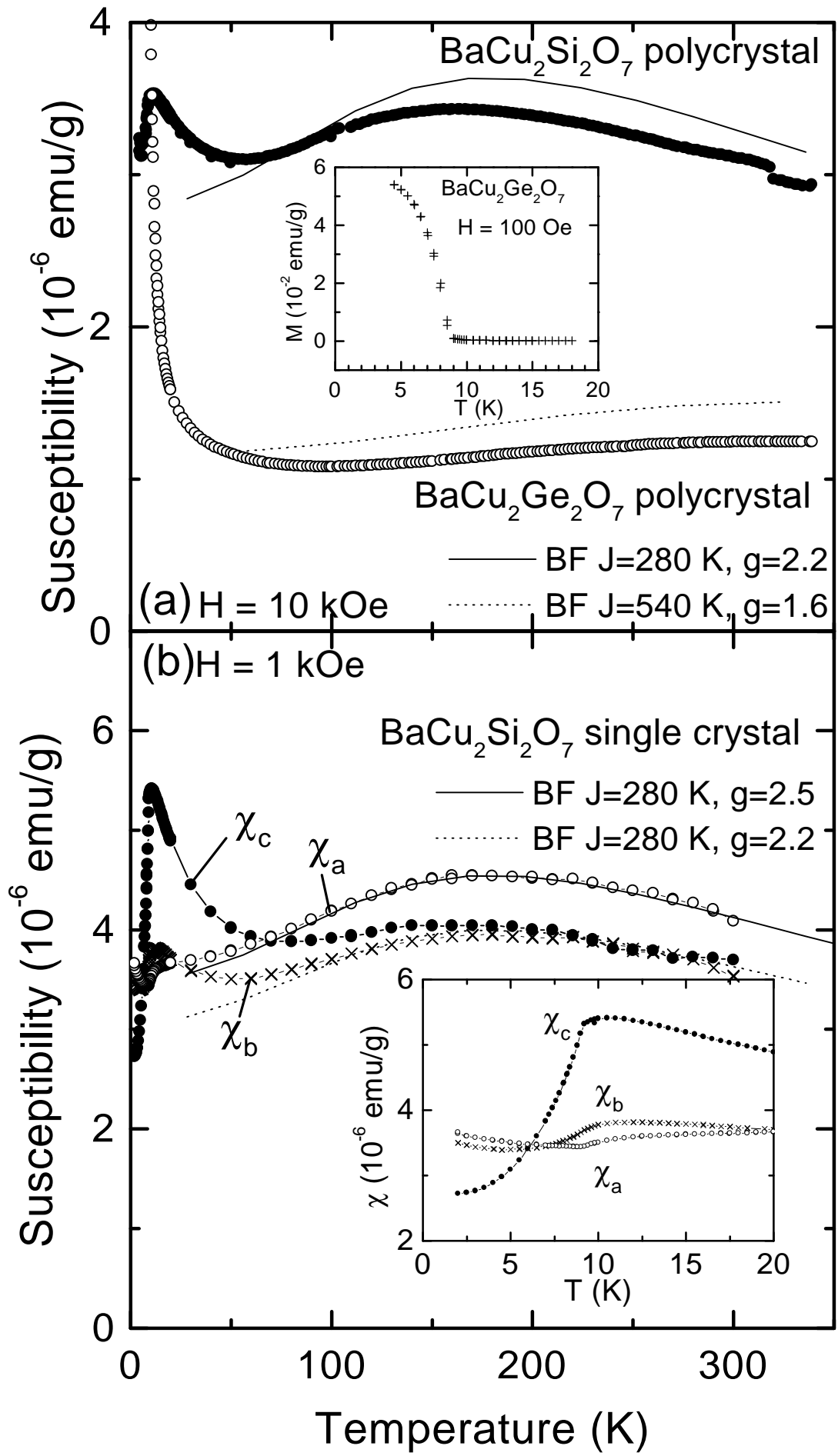
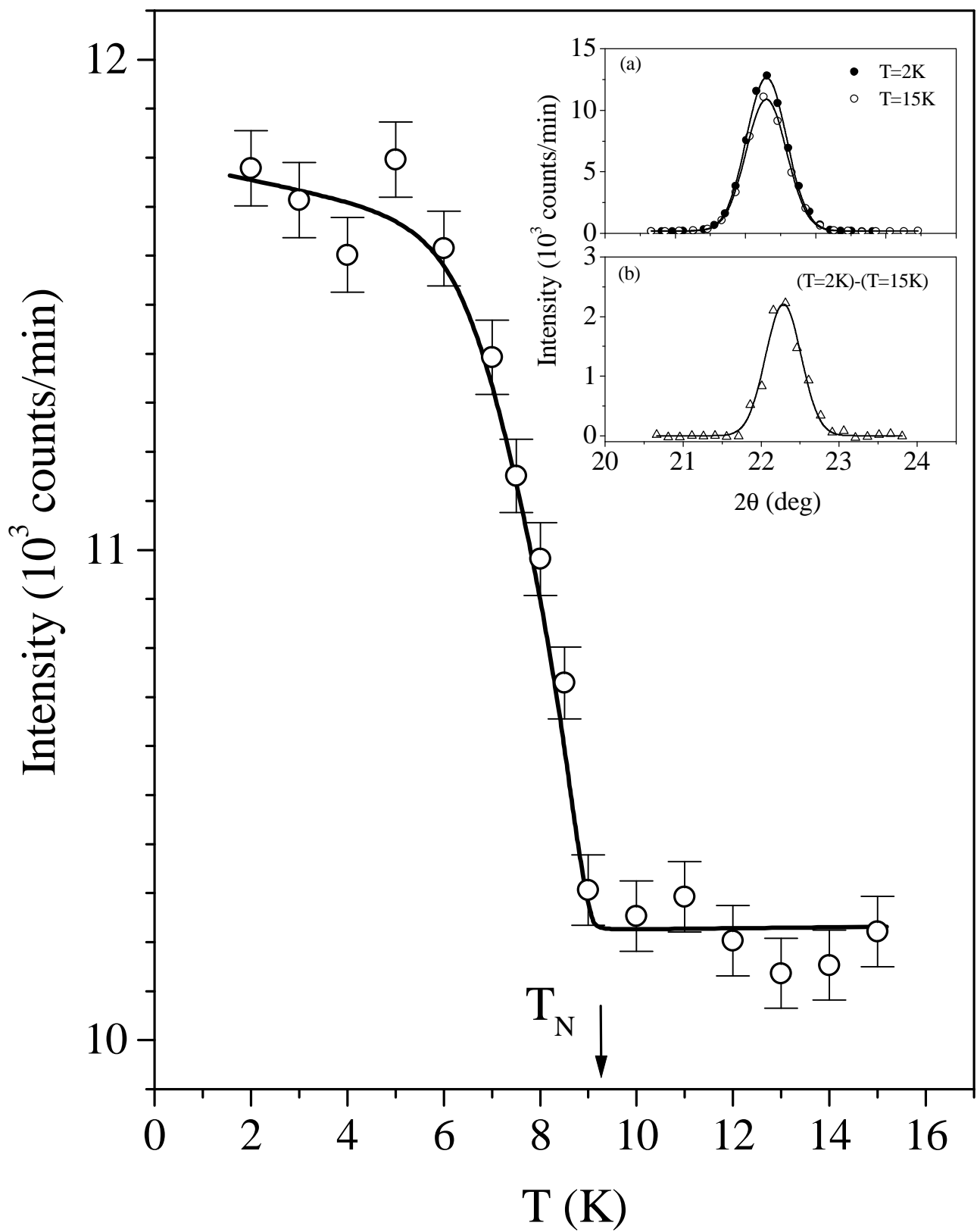
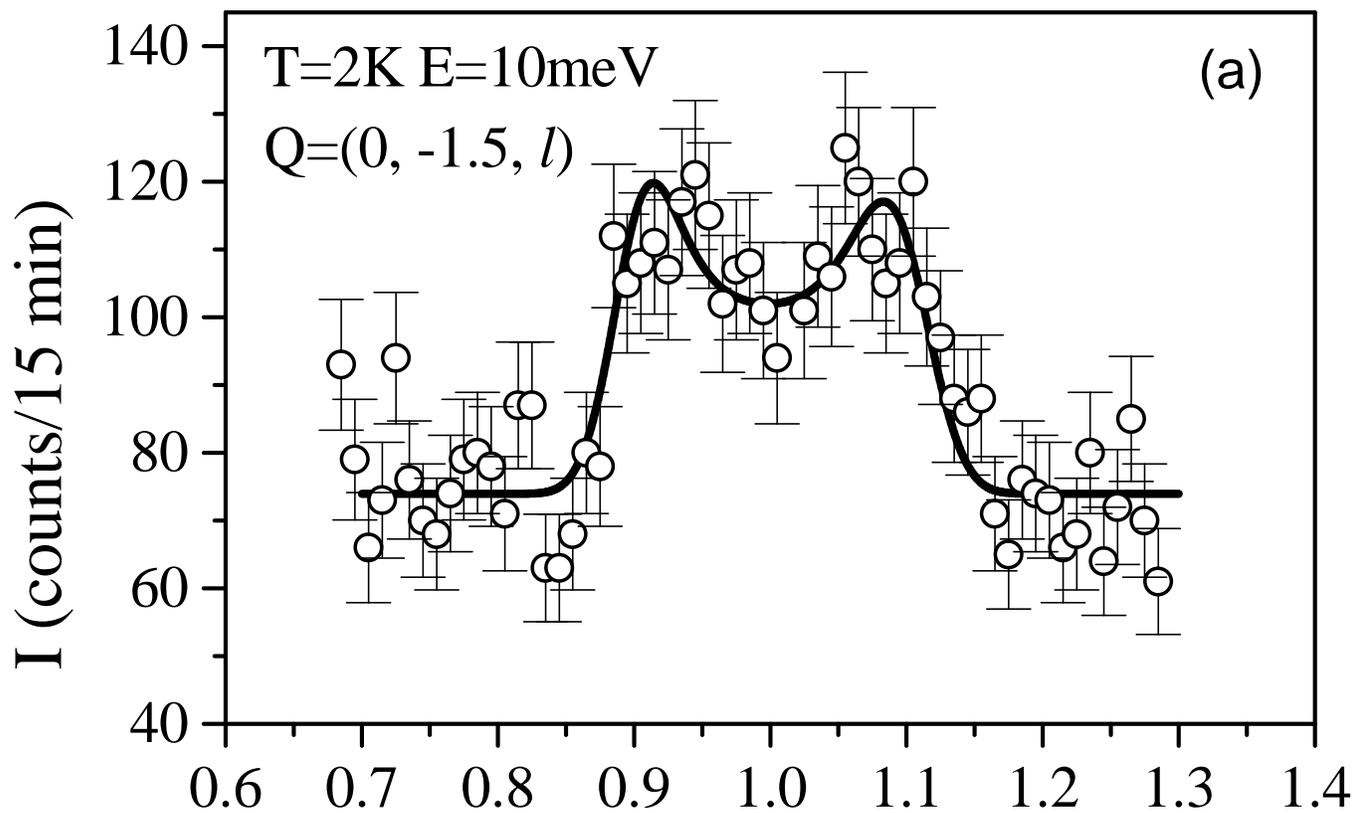
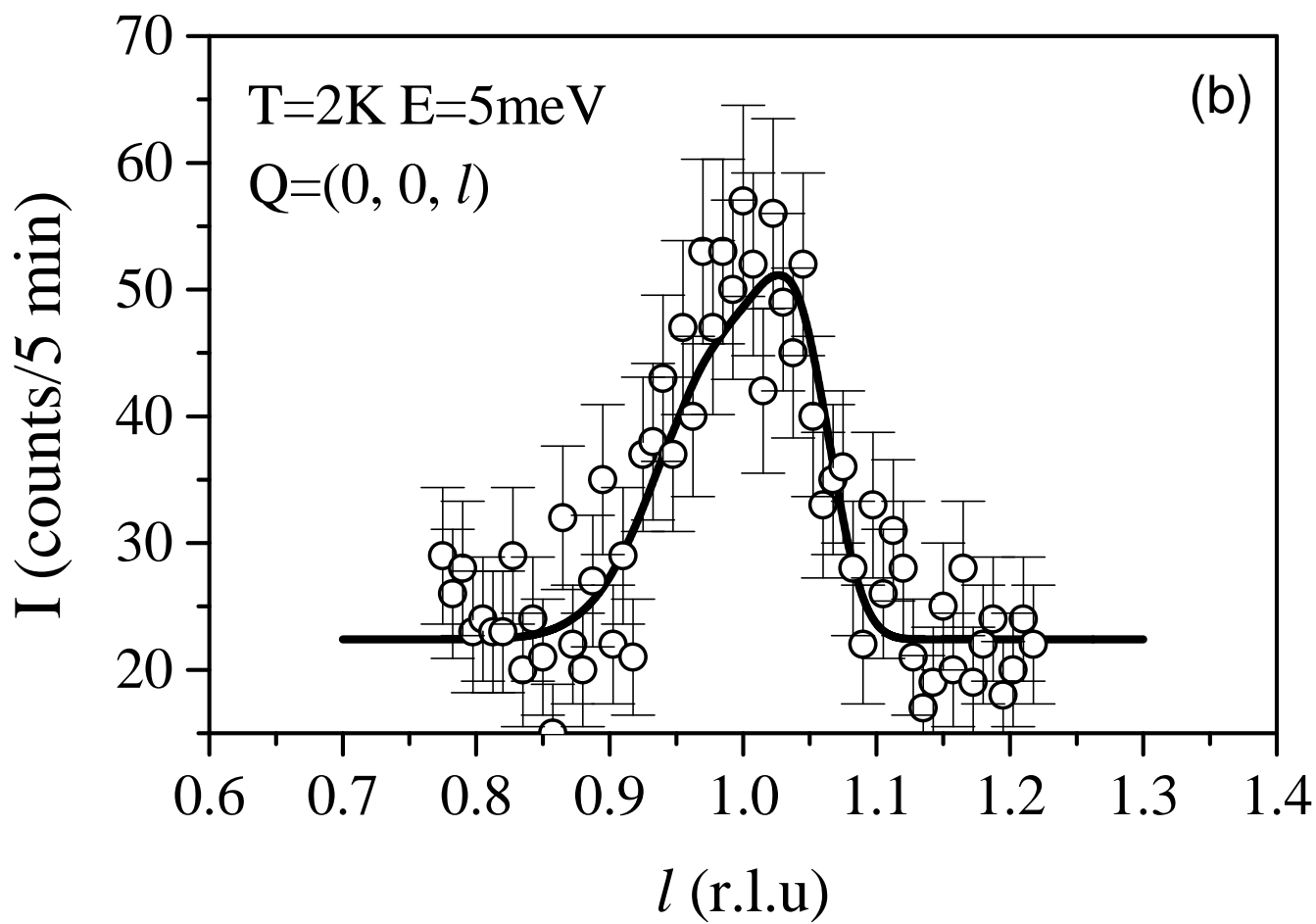
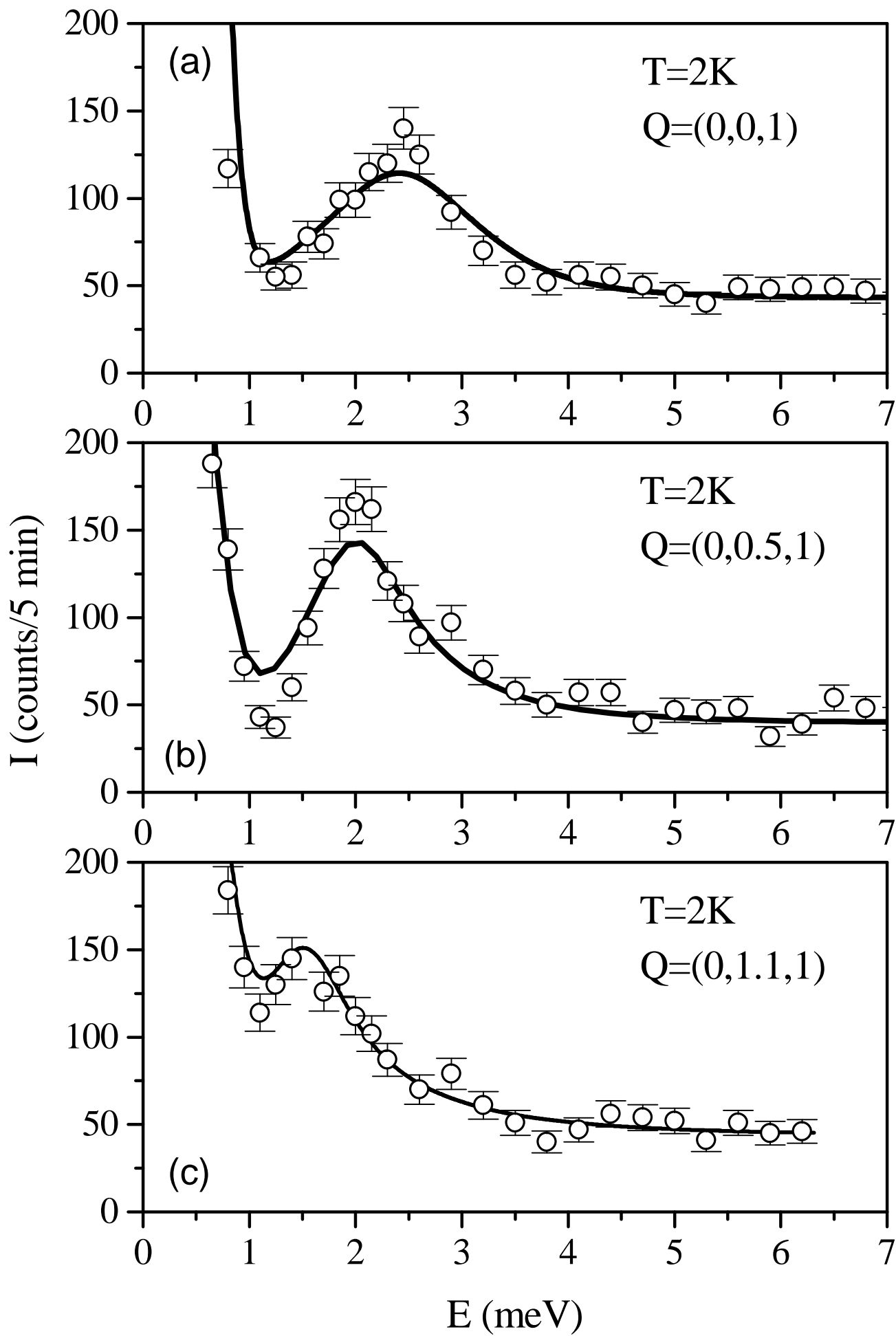
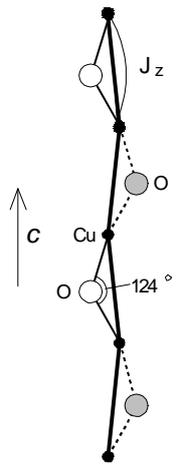
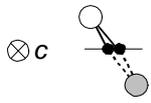
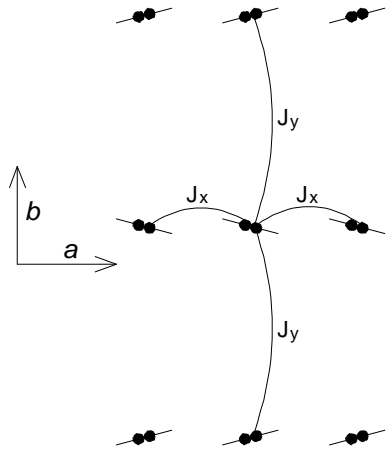
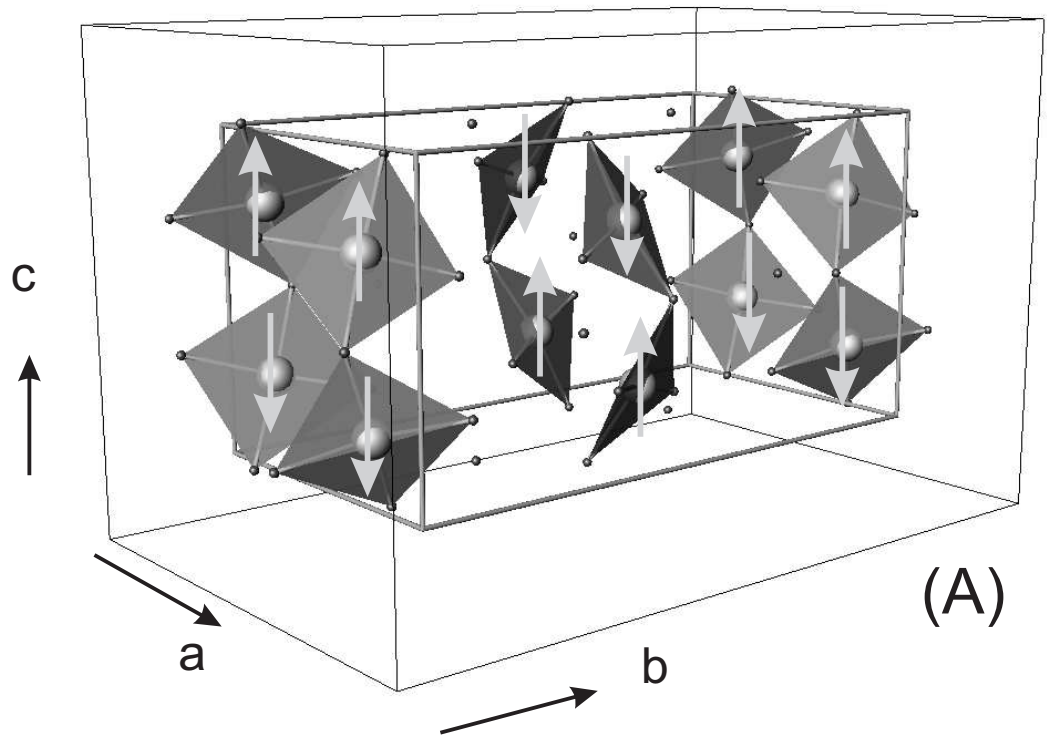
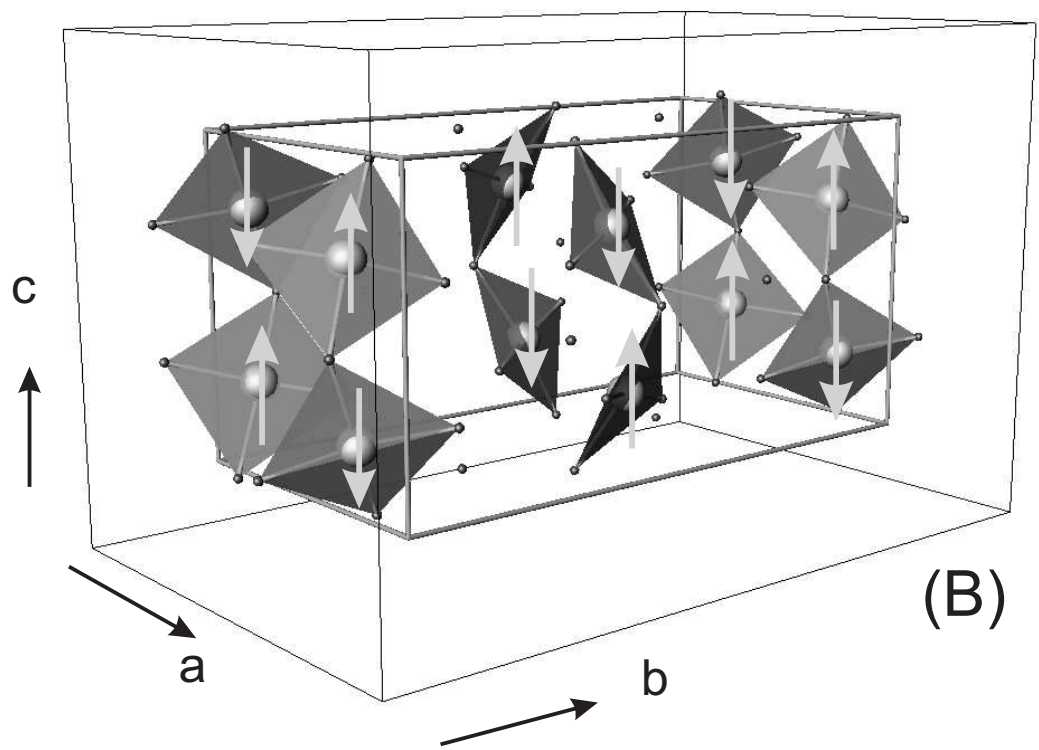
\end{document}